\documentstyle[aps,epsfig,multicol]{revtex}
\renewcommand{\narrowtext}{\begin{multicols}{2} \global\columnwidth20.5pc}
\renewcommand{\widetext}{\end{multicols}\global\columnwidth42.5pc}

\newcommand{\wideequationbegin}{
\widetext
\noindent
\begin{picture}(3.375,0)
  \put(0,0){\line(1,0){3.375}}
  \put(3.375,0){\line(0,1){0.08}}
\end{picture}}

\newcommand{\wideequationend}{
\hfill
\begin{picture}(3.375,0)
  \put(0,0){\line(1,0){3.375}}
  \put(0,0){\line(0,-1){0.08}}
\end{picture}
\narrowtext}

\begin{document}
\title{Angle dependence of Andreev scattering at semiconductor--superconductor
interfaces}
\draft
\author{Niels Asger Mortensen}
\address{Mikroelektronik Centret, Technical University of Denmark, Building 345 east, DK-2800 Lyngby, Denmark \&\\ Department of Physics, Technical University of Denmark, Building 309, DK-2800 Lyngby, Denmark}
\author{Karsten Flensberg}
\address{Danish Institute of Fundamental Metrology, Building 307, Anker Engelunds Vej 1, DK-2800 Lyngby, Denmark}
\author{Antti-Pekka Jauho}
\address{Mikroelektronik Centret, Technical University of Denmark, Building 345 east, DK-2800 Lyngby, Denmark}
\date{\today}
\maketitle

\begin{abstract}
We study the angle dependence of the Andreev scattering at a 
semiconductor--superconductor interface, generalizing the
one-dimensional theory of Blonder, Tinkham and Klapwijk. An increase
of the momentum parallel to the interface leads to suppression of the
probability of Andreev reflection and increase of the probability of 
normal reflection. We show that in the presence of a Fermi velocity 
mismatch between the semiconductor and the superconductor the angles
of incidence and transmission are related according to the well-known 
Snell's law in optics. As a consequence there is a critical angle of 
incidence above which only normal reflection exists. For two and 
three-dimensional interfaces a lower excess current compared to
ballistic transport with perpendicular incidence is found. Thus, the
one-dimensional BTK model overestimates the barrier strength for two and
three-dimensional interfaces.
\end{abstract}

\pacs{73.40.Gk, 74.90.+n, 85.25.-j}

\narrowtext

\section{Introduction}

An electron-like quasiparticle incident on a normal
conductor--superconductor (NS) interface from the normal side may become
Andreev reflected into a hole-like quasiparticle with reversal of the signs
of all three velocity components (retroflection) and of the energy (relative
to the Fermi level) as shown by Andreev \cite{ANDREEV64}. Later, Blonder,
Tinkham and Klapwijk (BTK) \cite{BTK82} calculated the scattering
probabilities at a NS interface within a model where the scattering at the
interface was represented by a delta-function potential barrier. The
calculations were based on the Bogoliubov--de Gennes (BdG) formalism \cite
{DEGENNES66}, for a one-dimensional (1D) geometry thus ignoring all effects
due to quasiparticles with a momentum parallel to the interface.

The BTK model has been widely used by experimentalists to model
normal-metal--superconductor junctions, and it has despite of its inherent
approximations been quite successful in describing the main features of these
devices. The quality of the junction interface has conveniently been
parametrized in terms of the normalized delta-function barrier strength.

A more complete theory was developed by Arnold \cite{ARNOLD} using
non-equilibrium Green's function techniques. The theory by Arnold
furthermore takes the three-dimensional (3D) nature of the interface into
account. However, the resulting expressions are complicated and require
substantial numerical work. Generalization of the BTK model to tunnel
barriers other than delta-function scattering potentials has been done by
Kupka in a number of papers \cite{KUPKA90,KUPKA97}. Recently, Kupka \cite
{KUPKA97} generalized the more realistic tunnel barrier model to include the
angel dependence of the scattering. He found that by treating the
scattering problem in the correct three dimensional picture, the effective
Andreev scattering is reduced and the normal scattering probability is
enhanced. Chaudhuri and Bagwell \cite{CHAUDHURI95} and De Raedt,
Michielsen and Klapwijk \cite{DERAEDT94} have also considered the angle dependence in their applications of the BdG formalism to the transport
  properties of NS interfaces. However, except for the 1D work of Blonder and Tinkham
\cite{BT83} the above mentioned papers all focused on the case where
there is no mismatch between densities (and hence Fermi wavelengths)
or between effective band masses of the two materials forming the NS
junction. In the case of
SNS junctions, Kupriyanov \cite{KUPRIYANOV} included effects of
the parallel degree of freedom and different Fermi velocities of the N
and S regions in his application of the Eilenberger equations to the
dc Josephson current in junctions with clean interfaces. Using the BdG
formalism, the effect of different Fermi velocities and effective masses was also considered by Sch\"{u}ssler and K\"{u}mmel \cite{SCHUSSLER93} and
Chrestin, Matsuyama and Merkt \cite{CHRESTIN94} in their numerical
studies of the dc Josephson current in Nb-InAs-Nb junctions.

Since much of the development in the recent years has been in structures
where superconductors are combined with semiconductors, the goal of this
paper is an analytical study of the importance of the different
quasiparticle propagation in the two materials, when the degrees of
freedom parallel to the interface and effects of the unequal Fermi
velocities and Fermi wavelengths are taken into account. This is
motivated by the observation that Andreev scattering cannot occur
above a critical angle where the momentum can no longer be
conserved. The critical angle depends on the ratio of the carrier
density of the semiconductor to the density of the superconductor. 
Therefore one may expect larger differences between 1D and 2D or 3D 
junctions for the case of a finite Fermi wave vector mismatch, which
is indeed what we find.

The effect of the angle dependence of the Andreev scattering probability is
however somewhat suppressed by the fact that the current is carried mostly
by particles incoming perpendicular to the interface in 2D or at angle
of $45^{\circ}$ in 3D. Therefore we suggest an experiment where the angle
dependence of Andreev scattering is probed in a more direct fashion, namely
a mesoscopic device which explores the ballistic motion of quasiparticles
and where the angle of incidence can be varied. Such a device is possible
due to the advances in fabrication of mesoscopic
semiconductor--superconductor interfaces (see e.g. \cite{AKAZAKI96}), which
have made it possible to study Andreev scattering in the ballistic regime.

The paper is organized as follows: In Sec. II the BdG formalism is
introduced, and in Sec. III the scattering probabilities at the interface is
calculated. These scattering probabilities are used in Sec. IV to calculate
current-voltage characteristics and related quantities. In Sec. V an
experiment is suggested and, finally, in Sec. VI discussions and conclusions
are given.

\section{The Bogoliubov--de Gennes formalism}

\label{sec:BdG}The BdG equations

\begin{equation}
\left( 
\begin{array}{cc}
\hat{{\cal H}}_{0}({\bf r}) & \Delta ({\bf r}) \\ 
\Delta ^*({\bf r}) & -\hat{{\cal H}}_{0}^*({\bf r})
\end{array}
\right) \psi ({\bf r})=E\psi ({\bf r})  \label{BdG}
\end{equation}
provide a microscopic formalism for studying inhomogeneous superconductors
and NS interfaces\cite{DEGENNES66}. Here, $\Delta ({\bf r})$ is the
superconducting order parameter and $\hat{{\cal H}}_{0}({\bf r})$ is the
Hamiltonian. In a general non-equilibrium situation, the Hamiltonian 
includes either a time-dependent vector potential or a spatially
dependent scalar potential. However, we follow BTK \cite{BTK82} and
neglect the effect of a finite bias on the scattering probabilities
which is justified if the height of the tunnel barrier is much higher
than the applied voltage and/or energy of the carriers (relative to
the Fermi level) \cite{KUPKA97}. For an interface where the position
of the conduction band and the effective mass change across the
interface, we use the effective mass approximation\cite{BASTARD,BURT}

\begin{equation}
\hat{{\cal H}}_{0}({\bf r})=-\hat{\nabla}\frac{\hbar ^{2}}{2m^\star({\bf r})}%
\hat{{\bf \nabla }}+U({\bf r})-\mu  \label{effectivemassH}
\end{equation}
where $U({\bf r})$ is total electrostatic potential, and $\mu $ is the
chemical potential. This approximation describes the spatial dependence of
the dispersion relation, and the form of the Hamiltonian ensures
conservation of the probability-current. For a discussion of justifications
of this approach, see Refs. \cite{BASTARD,BURT} and references
therein. We assume a parabolic dispersion so that the effective mass
$m^\star$ does not depend on energy (or momentum).

The solutions to Eq. (\ref{BdG}) are vectors in the so-called electron--hole
space (Nambu space), 
$\psi ({\bf r})=\left( u({\bf r}),v({\bf r})\right) ^{T}$, where 
$u({\bf r})$ is the electron-like quasiparticle amplitude satisfying
an ordinary electron-like Schr\"{o}dinger equation and $v({\bf r}) $
is the hole-like quasiparticle amplitude satisfying a time-reversed
Schr\"{o}dinger equation. In electron--hole space, a
probability-current-density can be associated with the wave function, and is
given by \cite{BTK82,DEGENNES66,BASTARD}

\begin{equation}
{\bf J}_{p}=\hbar{\rm Im}\left\{ u^{*}({\bf r})\frac{1}{m^\star({\bf r})}{\hat{{\bf \nabla }%
}}u({\bf r})-v^{*}({\bf r})\frac{1}{m^\star({\bf r})}{\hat{{\bf \nabla }}}v({\bf r})\right\} \;.
\label{Jp}
\end{equation}
The BdG equations and the conservation of the probability-current-density
form the basis for our treatment of scattering of quasiparticles at the NS
interface. Eq. (1) is used in calculating scattering amplitudes and
the corresponding scattering probabilities are found using Eq. (3).

\section{Scattering of quasiparticles at a NS interface}

We consider a planar NS interface lying in the $xy$-plane at $z=0$ with a
semi-infinite non-superconducting material for $z<0$ and a semi-infinite
superconductor for $z>0$. The superconducting order parameter is assumed to
vary in space only along the $z$-direction. In order to solve the BdG
equations, we include only scattering at the NS interface. Following BTK, we
model the scattering at the interface by a delta-function potential

\begin{equation}
U({\bf r})=H\delta (z)  \label{deltaapprox}
\end{equation}
where $H$ is the strength of the potential barrier. For simplicity we
neglect the phase of the pairing potential since only the absolute
value is important for the considered geometry. Furthermore, to avoid
self-consistent calculations, we take the superconducting order parameter to
be zero in the normal conductor and uniform in the superconductor, i.e.

\begin{equation}
\Delta ({\bf r})=\Delta _{0}\Theta (z),  \label{gapapprox}
\end{equation}
where $\Delta _{0}$ is the BCS value of the energy gap and $\Theta (z)$ is a
Heaviside function. Similarly for the effective masses of the two materials,
we assume that the mass changes abruptly across the interface 
\begin{equation}
m^\star({\bf r})=m^{\scriptscriptstyle {\rm (N)}}\Theta (-z)+m^{%
\scriptscriptstyle {\rm (S)}}\Theta (z)  \label{massapprox}
\end{equation}
where $m^{\scriptscriptstyle {\rm (N)}}$ and $m^{\scriptscriptstyle {\rm (S)}%
}$ are the effective masses of the normal conductor and the superconductor,
respectively. Eqs. (\ref{deltaapprox})-(\ref{massapprox}) represent the
simplest forms of $U({\bf r})$, $\Delta ({\bf r})$ and $m^\star({\bf r})$ still
capturing the main physics of the NS interface.

Due to the simple form of the NS barrier potential, the superconducting
order parameter and the effective mass, we can separate the variables and
express the solutions in the parallel direction as plane waves, i.e. 
$\psi ^{
\scriptscriptstyle {\rm (N,S)}}({\bf r})=\exp \left[ i\left( k_{x}^{%
\scriptscriptstyle {\rm (N,S)}}x+k_{y}^{\scriptscriptstyle {\rm (N,S)}%
}y\right) \right] \psi ^{\scriptscriptstyle {\rm (N,S)}}(z)$, where the
superscript $({\rm N},{\rm S})$ refers to the non-superconducting or
superconducting sides, respectively. Substituting this Ansatz into Eq. (\ref
{BdG}), yields the effective BdG equations for the $z$-direction
\wideequationbegin
\begin{equation}
\left( 
\begin{array}{cc}
\left[ -\frac{\hbar ^{2}}{2m^{\scriptscriptstyle {\rm (N,S)}}}\frac{%
\partial^2 }{\partial z^2}+H\delta (z)-\mu _{{\rm eff}}^{\scriptscriptstyle 
{\rm (N,S)}}\right] & \Delta_{0}\Theta (z) \\ 
\Delta _{0}\Theta (z) & -\left[ -\frac{\hbar ^{2}}{2m^{\scriptscriptstyle 
{\rm (N,S)}}}\frac{\partial^2 }{\partial z^2}+H\delta (z)-\mu _{{\rm eff}}^{%
\scriptscriptstyle {\rm (N,S)}}\right]
\end{array}
\right) \psi (z)=E\psi (z)  \label{BdG_eff}
\end{equation}
\wideequationend
\noindent where the effective chemical potential is defined as

\begin{equation}
\mu _{{\rm eff}}^{\scriptscriptstyle {\rm (N,S)}}\equiv\mu^{%
\scriptscriptstyle {\rm (N,S)}} -\frac{\hbar^{2}}{2m^{\scriptscriptstyle 
{\rm (N,S)}}}\left( \left[k_{x}^{\scriptscriptstyle {\rm (N,S)}%
}\right]^{2}+\left[k_{y}^{\scriptscriptstyle {\rm (N,S)}}\right]^{2}\right).
\label{mu_eff}
\end{equation}

Eq. (\ref{BdG_eff}) is mathematically identical to the 1D BdG equations
considered by BTK, and, therefore, we expect similar results for the
scattering probabilities. We adopt the notation of BTK
\cite[app. A]{BTK82} and all formulae for the eigenstates, scattering
states, wave vectors ($q^\pm$ and $k^\pm$) etc. are equivalent to
those of BTK, but with the important difference that the chemical
potential is replaced by an effective chemical potential which depends
on the parallel momentum according to Eq. (\ref{mu_eff}).

We follow BTK and consider an electron-like quasiparticle incident on
the NS interface from the normal side. At the interface it has an
amplitude $a$ of undergoing Andreev reflection, $b$ of normal
reflection, $c$ of normal transmission and $d$ of Andreev
transmission. The scattering amplitudes are obtained by matching the 
scattering states at the NS interface, using the appropriate boundary 
conditions for a delta-function potential barrier (see, e.g. 
\cite{BASTARD,GASIOROWICZ}). The matching results in following linear 
system determining $a$, $b$, $c$ and $d$
\wideequationbegin
\begin{equation}
\left( 
\begin{array}{cccc}
0 & 1 & -u_{0} & -v_{0} \\ 
1 & 0 & -v_{0} & -u_{0} \\ 
0 & \frac{2H}{\hbar ^{2}}-i\frac{q^{+}}{m^{\scriptscriptstyle {\rm (N)}}} & 
-i\frac{k^{+}}{m^{\scriptscriptstyle {\rm (S)}}}u_{0} & i\frac{k^{-}}{m^{%
\scriptscriptstyle {\rm (S)}}}v_{0} \\ 
\frac{2H}{\hbar ^{2}}+i\frac{q^{-}}{m^{\scriptscriptstyle {\rm (N)}}} & 0 & 
-i\frac{k^{+}}{m^{\scriptscriptstyle {\rm (S)}}}v_{0} & i\frac{k^{-}}{m^{%
\scriptscriptstyle {\rm (S)}}}u_{0}
\end{array}
\right) {\bf \cdot }\left( 
\begin{array}{c}
a \\ 
b \\ 
c \\ 
d
\end{array}
\right) =\left( 
\begin{array}{c}
-1 \\ 
0 \\ 
-\frac{2H}{\hbar ^{2}}-i\frac{q^{+}}{m^{\scriptscriptstyle {\rm (N)}}} \\ 
0
\end{array}
\right).  \label{abcd_eq}
\end{equation}
\wideequationend
Though complicated, the exact scattering probabilities can now be found and
numerical results for different values of the effective chemical potential
have been given by \u {S}ipr and Gy\"{o}rffy \cite{SIPRGYORFFY}. However,
the calculations may be simplified significantly for materials with a high
Fermi energy compared to the temperatures or bias voltages of interest. In
this limit transport only takes place near the Fermi level.

We choose polar coordinates and allow for a Fermi velocity mismatch by
considering a wave vector on the normal side given by ${\bf k}^{%
\scriptscriptstyle {\rm (N)}} = k_{{\rm F}}^{\scriptscriptstyle{\rm ( N)}%
}(\sin\theta\cos\phi,\sin\theta\sin\phi,\cos\theta)$ and a wave vector on
the superconducting side with $\left|{\bf k}^{\scriptscriptstyle{\rm (S)}}
\right| = k_{{\rm F}}^{\scriptscriptstyle{\rm (S)}}$.

The boundary conditions are satisfied only if 
$k_{x}^{\scriptscriptstyle{\rm 
(N)}}=k_{x}^{\scriptscriptstyle{\rm (S)}}$ and $k_{y}^{\scriptscriptstyle%
{\rm (N)}}=k_{y}^{\scriptscriptstyle{\rm (S)}}$, as dictated by the
translational invariance along the interface. This means that $k_{z}^{%
\scriptscriptstyle{\rm (S)}}=k_{{\rm F}}^{\scriptscriptstyle{\rm (S)}}\sqrt{%
\scriptstyle 1-r_{k}^{2}\sin ^{2}\theta }$, where the Fermi momentum ratio
is given by $r_{k}\equiv k_{{\rm F}}^{\scriptscriptstyle{\rm (N)}}/k_{{\rm F}%
}^{\scriptscriptstyle{\rm (S)}}$. The wave vectors on the
    normal conducting side can now
    be written as $q^\pm=k_{\rm F}^{\scriptscriptstyle\rm
      (N)}\sqrt{\cos^2\theta\pm E/\mu^{\scriptscriptstyle\rm (N)}}$
    and on the superconducting side we similarly get $k^\pm=k_{\rm F}^{\scriptscriptstyle\rm
      (S)}\sqrt{\left(1-r_k^2\sin^2\theta\right)\pm
      \sqrt{E^2-\Delta_0^2}/\mu^{\scriptscriptstyle\rm (S)}}$. This
    way of including the angle dependence is in accordance with
    Refs. \cite{KUPKA97,DERAEDT94,SCHUSSLER93,CHRESTIN94,SIPRGYORFFY,BEENAKKER95} but differs
    from the approach of Chaudhuri and Bagwell \cite{CHAUDHURI95} in
    which the angle corrections to the 1D expressions for $q^\pm$ and
    $k^\pm$ are approximated by $\cos\theta$ projection-factors. At low temperatures
$E/\mu^{\scriptscriptstyle\rm (N)} \sim \sqrt{%
\scriptstyle E^{2}-\Delta _{0}^{2}}/\mu^{\scriptscriptstyle\rm (S)} \ll 1$, and therefore, we
apply the Andreev approximation: $%
k^{+}=k^{-}=k_{z}^{\scriptscriptstyle{\rm (S)}}$ and $q^{+}=q^{-}=k_{z}^{%
\scriptscriptstyle{\rm (N)}}$. For values of
$\mu^{\scriptscriptstyle\rm (N,S)}$ relevant for normal
metals and low-temperature superconductors this approximation is valid for
angles of incidence $\theta \,{\scriptstyle \stackrel{<}{\sim }}\,\pi /2$.
Semiconductors have much lower Fermi energies as compared to those of normal
metals, but, even for $T/T_{{\rm F}}\sim 1/50$ the approximation is
reasonable. For angles in the vicinity of $\theta \sim \pi /2$ the
approximation becomes inaccurate. However, quasiparticles with vanishing
perpendicular momentum do not contribute significantly to the perpendicular
current and their effect in the IV curves and related quantities will thus
not be important for semiconductors either. We have in fact checked
this for the excess current by numerically solving Eq. (16) and found
less than a half
\newpage
\widetext
\begin{table}[tbp]
\caption{Scattering probabilities at NS interface as a function of the
normalized excitation energy $\tilde{E}\equiv E/\Delta_0$.}
\label{ABCD}
\begin{tabular}{cccc}
& $|\tilde{E}| < 1$, $\theta<\theta_c$ & $|\tilde{E}| > 1$, $\theta<\theta_c$
& $\theta \geq \theta_c$ \\ \hline
\vspace{-2mm} &  &  &  \\ 
$A(\tilde{E},\theta)$ & $\frac{1}{\tilde{E}^2 + \left(1 - \tilde{E}%
^2\right)\left(1 + 2 Z_{{\rm eff}}^2(\theta)\right)^2}$ & $\frac{1}{\left[|%
\tilde{E}| + \sqrt{\tilde{E}^2 - 1}\left(1 + 2 Z_{{\rm eff}%
}^2(\theta)\right)\right]^2}$ & $0$\vspace{2mm} \\ 
$B(\tilde{E},\theta)$ & $1-A(\tilde{E},\theta)$ & $\frac{4\left(\tilde{E}^2
- 1\right)\left(Z_{{\rm eff}}^4(\theta) + Z_{{\rm eff}}^2(\theta)\right)}{%
\left[|\tilde{E}| + \sqrt{\tilde{E}^2 - 1}\left(1 + 2 Z_{{\rm eff}%
}^2(\theta)\right)\right]^2}$ & $1$ \vspace{2mm} \\ 
$C(\tilde{E},\theta)$ & $0$ & $\frac{2\sqrt{\tilde{E}^2-1}\left(|\tilde{E}|
+ \sqrt{\tilde{E}^2 -1}\right)\left(1 + Z_{{\rm eff}}^2(\theta)\right)}{%
\left[|\tilde{E}| + \sqrt{\tilde{E}^2 - 1}\left(1 + 2 Z_{{\rm eff}%
}^2(\theta)\right)\right]^2}$ & $0$\vspace{2mm} \\ 
$D(\tilde{E},\theta)$ & $0$ & $\frac{2\sqrt{\tilde{E}^2-1}\left(|\tilde{E}|
- \sqrt{\tilde{E}^2 -1}\right) Z_{{\rm eff}}^2(\theta)}{\left[|\tilde{E}| + 
\sqrt{\tilde{E}^2 - 1}\left(1 + 2 Z_{{\rm eff}}^2(\theta)\right)\right]^2}$
& $0$%
\end{tabular}
\end{table}
\narrowtext 

\noindent percent deviations for the GaAs 2DEG considered in
FIG. 2. With these approximations the amplitudes become

\begin{eqnarray}
a &=&\frac{u_{0}v_{0}}{\gamma }  \label{aeff_amplitude} \\
b &=&-\frac{\left( u_{0}^{2}-v_{0}^{2}\right) \left( \Gamma \left( \frac{Z}{%
\cos \theta }\right) ^{2}+\frac{1-\Gamma ^{2}r^{2}}{4\Gamma r}+i\Gamma \sqrt{%
r}\frac{Z}{\cos \theta }\right) }{\gamma }  \label{beff_amplitude} \\
c &=&\frac{u_{0}\left( \frac{1+\Gamma r}{2}-i\Gamma \sqrt{r}\frac{Z}{\cos
\theta }\right) }{\gamma }  \label{ceff_amplitude} \\
d &=&\frac{iv_{0}\left( \Gamma \sqrt{r}\frac{Z}{\cos \theta }-i\frac{%
1-\Gamma r}{2}\right) }{\gamma }\;,  \label{deff_amplitude}
\end{eqnarray}
where we define $\gamma \equiv u_{0}^{2}+\left( u_{0}^{2}-v_{0}^{2}\right)
Z_{{\rm eff}}^{2}(\theta )$, and where

\begin{equation}
Z_{{\rm eff}}(\theta )=\sqrt{\Gamma (\theta )\left( \frac{Z}{\cos \theta }%
\right) ^{2}+\frac{(\Gamma (\theta )r_{v}-1)^{2}}{4\Gamma (\theta )r_{v}}}
\label{Z_eff}
\end{equation}
is an effective barrier strength, $r_{v}\equiv v_{{\rm F}}^{%
\scriptscriptstyle {\rm (N)}}/v_{{\rm F}}^{\scriptscriptstyle {\rm (S)}}$ is
the Fermi velocity ratio and $\Gamma (\theta )\equiv \cos \theta /\sqrt{%
\scriptstyle 1-r_{k}^{2}\sin ^{2}\theta }$. The dimensionless barrier
strength $Z\equiv H/\hbar \sqrt{\scriptstyle v_{{\rm F}}^{{\rm (N)}}v_{{\rm F%
}}^{{\rm (S)}}}$ was introduced by Blonder and Tinkham \cite{BT83}.

In order to obtain the scattering probabilities $A$, $B$, $C$ and $D$ we use
the conservation of the probability-current-density, Eq. (\ref{Jp}). For the $z$-direction
this yields
\wideequationbegin
\begin{equation}
1 = \underbrace{|a|^2}_{A}+ \underbrace{|b|^2}_{B}+ \underbrace{%
\Theta(|E|-\Delta_0)\frac{|u_0|^2 - |v_0|^2}{\Gamma r}|c|^2}_{C}+ 
\underbrace{ \Theta(|E|-\Delta_0)\frac{ |u_0|^2 - |v_0|^2}{\Gamma r}|d|^2}%
_{D}.
\end{equation}
\wideequationend

It turns out that the scattering probabilities of the BTK model can still be
applied provided that the dimensionless barrier strength is replaced by the
introduced effective barrier strength given in Eq. (\ref{Z_eff}). For
perpendicular incidence ($\theta =0$) this result agrees with the BTK result
($r_{v}=1$) \cite{BTK82} and the Blonder--Tinkham result \cite{BT83} which
includes the possibility of a Fermi velocity mismatch. For a general angle
of incidence and matching Fermi velocities and Fermi momenta 
($r_{v}=r_{k}=1$) the result reduces to that obtained by Kupka 
\cite{KUPKA97}.

As mentioned, the wave vectors of the transmitted waves have the form $k_z^{%
\scriptscriptstyle{\rm (S)}} = k_{{\rm F}}^{\scriptscriptstyle{\rm (S)}} 
\sqrt{\scriptstyle 1 - r_k^2\sin^2\theta}$. The square root defines a
critical angel of incidence $\theta_c$, above which the solutions are
evanescent and below which we have propagating waves, i.e. $\theta_c =
\arcsin(1/r_k)$ for $r_k>1$. The
physical reason for the critical angle is that the parallel momentum exceeds
the Fermi momentum of the superconductor and thus momentum can not be
conserved. For $r_k \leq 1$ there is no critical angle due to the
parallel momentum not being conserved. Going beyond the Andreev
approximation introduces another energy dependent critical angle
$\tilde{\theta}_c = \arcsin \sqrt{1-E/\mu^{\scriptscriptstyle \rm
    (N)}}$ caused by the wave vector $q^-$ of the Andreev scattering
state being imaginary \cite{SIPRGYORFFY}. However, the later critical
angle has little consequences for our results for the same reasons as
when we discussed the validity of the Andreev approximation. In
Sec. \ref{experiment}, we suggest how the angle dependence may be probed.

\begin{figure}[tbp]
\begin{center}
\epsfig{file=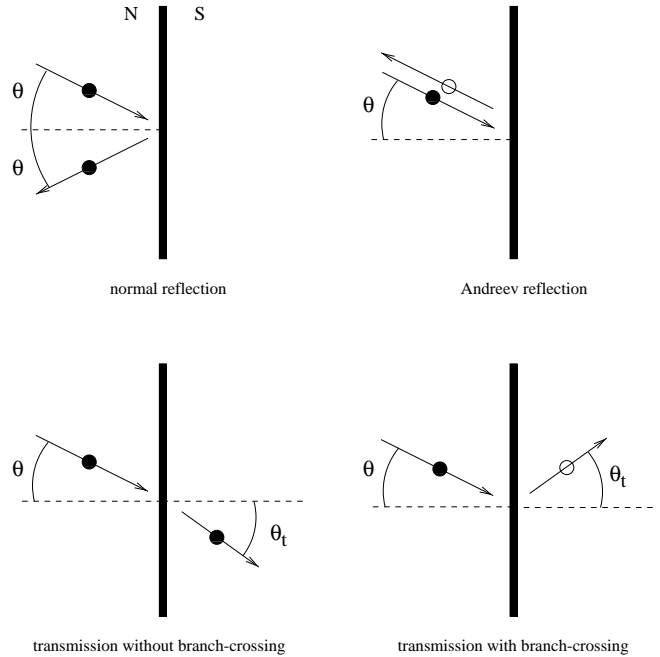, width=0.99\columnwidth,clip}
\end{center}
\caption{Directions of the transmitted and reflected waves in real space.
The full dots ({\protect\large $\bullet$}) represent quasiparticles of
predominantly electron-like character and the open dots 
({\protect\large $\circ$}) represent quasiparticles of predominantly 
hole-like character.}
\label{direction}
\end{figure}

The directions of the reflected and transmitted waves can be obtained by
considering the probability-current-density and the result is sketched in
Fig.\ \ref{direction}. The angle of reflection $\theta_r$ coincides with the
angle of incidence $\theta$ and the angle of transmission is given by

\begin{equation}
\sin \theta _{t}=r_{k}\sin \theta
\end{equation}
in analogy with Snell's law in optics as it was also found by
Kupriyanov \cite{KUPRIYANOV}. The general results for the
scattering probabilities including the possibility of a Fermi velocity
mismatch are summarized in Table \ref{ABCD}. We conclude that the scattering
probabilities of the BTK model still apply, provided that the dimensionless
barrier strength is replaced by an angle dependent effective barrier
strength.

As the angle of incidence is increased, we observe an increasing effective
barrier strength and therefore Andreev reflection is suppressed when the
parallel momentum becomes much larger than the perpendicular momentum \cite{KUPKA97}. In
the same way, normal reflection increases when the parallel momentum
increases. Nevertheless, we still have unit probability for Andreev
reflection at the gap edge for all angles of incidence ($\theta < \theta_c$)
and for all barrier strengths.

\section{Current-voltage characteristic, excess current and differential
conductance}

We calculate the current on the normal side of the interface where the
current is carried only by single quasiparticles and no supercurrent. The
current density in the $z$-direction is given by

\begin{equation}
J_{z}=\sum_{\sigma }\int \frac{{\rm d}^{d}{\bf k}}{(2\pi )^{d}}e{\bf v}\cdot 
\hat{{\bf e}}_{z}f^{\scriptscriptstyle {\rm (N)}}({\bf k})
\end{equation}
where $d=1,2,3$ is the dimension of the electron-gas and $f^{%
\scriptscriptstyle {\rm (N)}}({\bf k})$ is the non-equilibrium distribution
function on the normal side of the interface. This approach neglects
coherent effects of the propagation of electron-like and hole-like
quasiparticles in the normal region and it applies to NS interfaces
with a ballistic normal region and/or NS interfaces where the length
of the normal region is large on the scale of the phase coherence
length. The integration is performed using polar coordinates
appropriate for a 1D electron gas, a two-dimensional (2D) electron
gas, and a 3D electron gas, respectively. The 1D case corresponds to the BTK model.

In general the non-equilibrium distribution function can be found from a
suitable transport equation, e.g. Boltzmann equation. Instead of taking this
path, we follow BTK and assume that all quasiparticles incident from the
reservoir are distributed in accordance with the Fermi--Dirac equilibrium
distribution function with a shift in energy due to the applied voltage.
When current flows, the reservoir is not in true equilibrium. However, the
voltage drop across the normal region can be accounted for by an Ohmic
series resistance. We calculate $f^{\scriptscriptstyle {\rm (N)}%
}(E,\theta)=\Theta(\pi/2-\theta)f_\rightarrow^{\scriptscriptstyle {\rm (N)}%
}(E,\theta) +\Theta(\theta-\pi/2)f_\leftarrow^{\scriptscriptstyle {\rm (N)}%
}(E,\theta)$ by considering the two sub-populations separately. If we take
the chemical potential of the superconductor as reference, we get $%
f_\rightarrow^{\scriptscriptstyle{\rm (N)}}(E,\theta,V) = f_0(E-eV)$ for the
sub-population of quasiparticles with a positive momentum in the $z$%
-direction. The sub-population of quasiparticles with a negative momentum is
\wideequationbegin
\begin{eqnarray}
f_\leftarrow^{\scriptscriptstyle{\rm (N)}}(E,\pi-\theta,V) &=&
A(-E,\theta)\left[1-f_\rightarrow^{\scriptscriptstyle{\rm (N)}
}(-E,\theta,V)\right] 
+ B(E,\theta)f_\rightarrow^{\scriptscriptstyle{\rm (N)}}(E,\theta,V)  
\nonumber \\
&&\quad + C(E,\theta)f_\leftarrow^{\scriptscriptstyle{\rm (S)}
}(E,\theta_S,V) 
+ D(E,\theta) f_\leftarrow^{\scriptscriptstyle{\rm (S)}
}(E,\theta_S,V).
\end{eqnarray}
\wideequationend
\noindent Here the first term represents Andreev reflection of time-reversed
quasiparticles, the second term represents normal reflection, and the last
two terms represent transmission of quasiparticles from the superconductor
where $f_\leftarrow^{\scriptscriptstyle{\rm (S)}}(E,\theta_S,V)=f_0(E)$ and 
$\theta_S = \arcsin\left(r_k \sin \theta\right)$. Using the translational
invariance along the interface, the sum rule $1=A+B+C+D$, and the symmetries
with respect to energy yields the normalized current

\begin{equation}
I=\frac{\Delta _{0}}{eR_{{\rm N}}}\int_{-\infty }^{\infty }
{\rm d}\tilde{E}\,
\overline{T}(\tilde{E})
\left[ f_{0}(\tilde{E}-eV/\Delta _{0})-f_{0}(\tilde{E})\right],
\end{equation}

\widetext
\begin{figure}[tbp]
\begin{center}
\epsfig{file=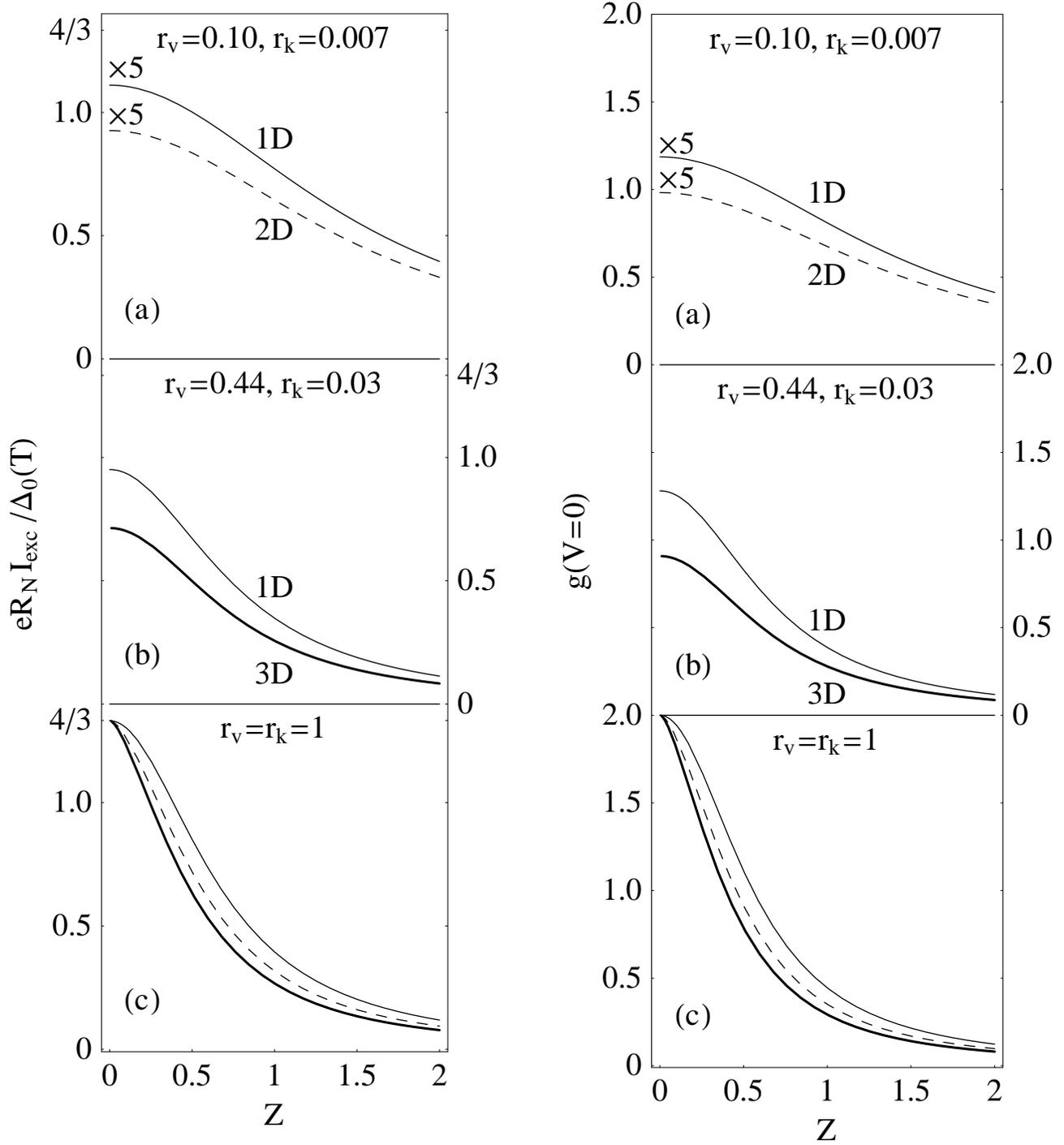 , width=0.99\columnwidth,clip}
\end{center}
\caption{Left panel: Normalized excess current $eR_{{\rm N}}I_{{\rm exc}%
}/\Delta_0(T)$ as a function of the dimensionless barrier strength $Z$.
Right panel: Normalized differential conductance at zero bias $g(V=0)$ as a
function of the dimensionless barrier strength $Z$. The thin lines, dashed
lines and thick lines correspond to the 1D BTK model, the 2D and 3D
calculations, respectively. The temperature dependence of the excess current
is entirely contained within the normalization through $\Delta_0(T)$ and the
differential conductance is plotted at $T=0\,{\rm K}$. The excess current
and the zero bias differential conductance are shown for (a) parameters
appropriate for a 2D GaAs-Al interface with a Fermi temperature $T_{{\rm F}}
\simeq 100\, {\rm K}$ of GaAs, (b) parameters appropriate for the 3D GaAs-Al
interface studied by Taboryski {\em et al.} \protect\cite{TABORYSKI96} and
(c) matching Fermi velocities and Fermi momenta. For the shown 2D result in
(a) and 3D result in (b) the overestimation of $Z$ for a perfect interface
is $0.68$ and $0.45$, respectively.}
\label{excess}
\end{figure}
\narrowtext
\noindent where $\tilde{E}\equiv E/\Delta _{0}$ is the normalized excitation
energy, $R_{{\rm N}}$ is the normal state resistance and

\begin{eqnarray}
\overline{T}_{\scriptscriptstyle {\rm 1D}}(\tilde{E}) &\equiv &
\left[ 1+A(\tilde{E},0)-B(\tilde{E},0)\right], \\
\overline{T}_{\scriptscriptstyle{\rm 2D}}(\tilde{E}) &\equiv &\int_{0}^{\pi
/2}\hspace{-4mm}{\rm d}\theta \,\frac{\cos \theta }{2\pi }\left[ 1+A(\tilde{E%
},\theta )-B(\tilde{E},\theta )\right], \\
\overline{T}_{\scriptscriptstyle {\rm 3D}}(\tilde{E}) &\equiv &\int_{0}^{\pi
/2}\hspace{-4mm}{\rm d}\theta \,\frac{\sin \theta \cos \theta }{2}\left[ 1+A(%
\tilde{E},\theta )-B(\tilde{E},\theta )\right],\hspace{-1mm}
\end{eqnarray}
are effective transmission coefficients for electrical current. The current
for two and three-dimension systems have the same qualitative form as in the
1D BTK model. However, quantitative changes are seen in the excess current
and the differential conductance.

The effective transmission coefficients are in general larger than the
corresponding normal state transmission coefficients and this effect gives
rise to a voltage dependent excess current compared to the normal state,
where often the high-voltage limit

\begin{equation}
I_{{\rm exc}}\equiv \lim_{eV\gg \Delta _{0}}\left[ I(V)-\lim_{\Delta
_{0}\rightarrow 0}I(V)\right]
\end{equation}
is of interest from an experimental point of view.

We have shown above, that a large parallel momentum suppresses the Andreev
reflection probability and thus we expect to see a lower excess current in
the three or two-dimensional limit as compared to the case of perpendicular
incidence. This is seen in the left panel of Fig.\ \ref{excess}. For perfect
($Z=0$) 2D and 3D interfaces with non-matching Fermi velocities and Fermi
momenta, the 1D BTK model overestimates the barrier strength $Z$
significantly. For the shown 2D and 3D results the overestimation of $Z$ for
a perfect interface is $0.68$ and $0.45$, respectively.

At low temperatures the normalized differential conductance, 
$g\equiv G_{{\rm NS}}/G_{{\rm NN}}$, is given by 
\begin{equation}
g(V)=\overline{T}(eV/\Delta _{0})/\overline{T}_{{\rm N}}
\end{equation}
where $\overline{T}_{{\rm N}}$ is the effective transmission probability
when the superconductor is in the normal state. In the right panel of 
Fig. \ref{excess}, results at zero bias are shown. Similarly to the excess
current, we find a lower zero-bias conductance with raising dimensionality
as compared to the case of transport with perpendicular incidence.

As an application of the present results, we now consider recent experiments
by Taboryski {\it et al.} \cite{TABORYSKI96} who reported on Andreev
reflections at interfaces between GaAs (3DEG) and superconducting Al films.
The material parameters are $r_v \simeq 0.44$ and $r_k \simeq 0.03$ and from
the excess current of the 1D BTK model, Taboryski {\it et al.} deduce the
dimensionless barrier strength $Z_{{\rm fit}}$ to fall in the range from $0.7
$ to $0.9$. Comparing with (b) in the left panel of Fig.\ \ref{excess} we
find the barrier strength to fall in the range from $0.5$ to $0.7$. For 
GaAs, the energy dependence of the effective mass due to non-parabolicity 
is negligible within $\sim 50 \Delta_{\rm Al}$ of the Fermi level 
\cite{ADACHI93} for both the cases considered in FIG. 2. Since the 
zero-bias conductance is a Fermi-surface property no restrictions have
been made by neglecting the energy dependence of the effective
mass. For the high voltage limit of the excess current, the corrections 
due to a non-parabolic conduction band are small. For InAs, as considered 
by Sch\"{u}ssler and K\"{u}mmel \cite{SCHUSSLER93} the energy dependence 
due to non-parabolicity is more pronounced.

\section{Suggested experiment}

\label{experiment}

\begin{figure}[tbp]
\begin{center}
\epsfig{file=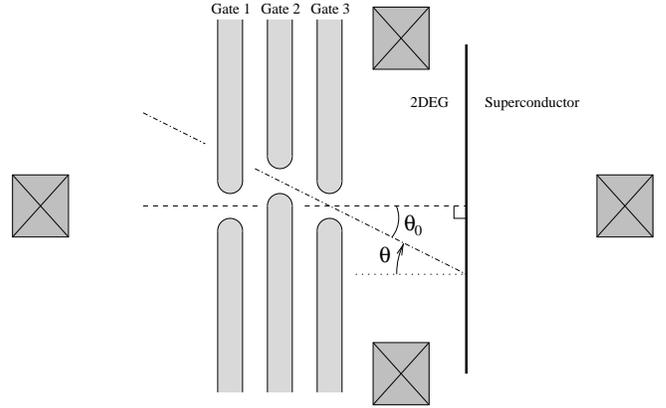 , width=0.99\columnwidth,clip}
\end{center}
\caption{Interface between a ballistic 2DEG and a superconductor. The three
gates may individually be applied a gate-voltage and in this way it is
possible to have perpendicular incidence or a finite angle of incidence
defined by for example gate 2 and gate 3.}
\label{2DEG-S_gate}
\end{figure}

Benistant {\it et al.}
  \cite{BENISTANT85} have studied the angle dependence of Andreev
  scattering at Ag-Pb interfaces experimentally by using a magnetic
  focusing technique. The quasiparticles are injected to a very
  clean (ballistic) 3D Ag crystal through a point contact and the angle
  of incidence at the NS interface is controlled by a weak magnetic
  field. We suggest a variant based on an interface between a
  ballistic two-dimensional electron-gas (2DEG) and a superconductor
  and the technological opportunity of defining the angle of incidence geometrically. By applying gates on top of the
2DEG, it is possible to control the angle of incidence as sketched in 
Fig. \ref{2DEG-S_gate}. 

If gate 1 and gate 3 are both negatively biased 
whereas gate 2 is turned off then quasiparticles have perpendicular 
incidence. However, biasing gate 2 and gate 3, while gate 1 is turned off, 
a finite angle of incidence can be achieved.

The angle dependence can be studied by measuring the differential
conductance. At low temperatures the normalized differential conductance, 
$g(eV,\theta)\equiv G_{{\rm NS}}(eV,\theta)/G_{{\rm NN}}(eV,\theta)$, is
given by

\begin{equation}
g(eV,\theta )=\frac{1+T(\theta )A(eV,\theta )-T(-\theta )B(eV,\theta )}{%
1-T(-\theta )\tilde{B}(\theta )}  \label{normalized_g}
\end{equation}
where $\tilde{B}=Z_{{\rm eff}}^{2}/(1+Z_{{\rm eff}}^{2})$ is the normal
reflection probability when the superconductor is in the normal state. The
transmission for quasiparticles leaving the interface through the gates, 
$T(\theta )$, is peaked around $\theta =\theta _{0}$, with a width depending
on the geometry. The possibility of multiple Andreev reflections can be
neglected if the phase-relaxation length is less than four times the
distance between the interface and the gate nearest to the interface.

The experimental curves for $g(eV,0)$ and $g(eV,\theta_0)$ may be fitted to
Eq. (\ref{normalized_g}) with the transmission, $T(\theta_0)$, and the
barrier strength, $Z$, as fitting-parameters. Further information on $Z$ and 
$T(\theta_0)$ may be obtained from the normal state conductance. When the
superconductor is in the normal state the Landauer formula yields the
conductance

\begin{equation}
G_{{\rm NN}}(\theta _{0})=\frac{2e^{2}}{h}MT(\theta _{0})\left[ 1-\tilde{B}%
(\theta _{0})\right]
\end{equation}
where $M$ is the number of modes. Thus, agreement with the angle dependence
is found if fits of the experimental curves for $g(eV,0)$ and $g(eV,\theta
_{0})$ can be obtained using the same $Z$-value.

\section{Discussion and conclusion}

The angle dependence of scattering at NS interfaces is of important
consequence when the NS interface has two or three-dimensional nature. The
parallel degrees of freedom also have important consequences in
superconducting mesoscopic transport. If the current is carried by more than
a single mode different modes represent different momenta parallel to the
interface, and thus the scattering amplitudes depend on the mode index. More
details may be found in a review of scattering theory in mesoscopic NS
structures by Beenakker \cite{BEENAKKER95}.

We have investigated the angle dependence of scattering of quasiparticles at
NS interfaces using the framework of Bogoliubov--de Gennes. As a main result
the scattering probabilities of the BTK model may still be applied provided
that the scattering strength is replaced by an effective angle dependent
barrier strength. This modified effective scattering parameter agrees with
previous calculations of BTK \cite{BTK82}, Blonder and Tinkham \cite{BT83}
and Kupka \cite{KUPKA97}. One of the consequences is that the Andreev
reflection is suppressed for large angles of incidence and the normal
reflection is increased towards unity. In the presence of a Fermi momentum
mismatch, we find the angles of incidence and transmission to be related and
in analogy with Snell's law, we find that above a certain critical angle of
incidence we only have normal reflection.

Furthermore, the results of the angle dependence have been applied to a NS
interfaces with one, two and three-dimensional nature where we find that the
1D BTK model overestimates the barrier strength. Calculations show that for
certain material parameters and clean interfaces the corrections may be
significant. However, the over-all qualitative predictions of the 1D BTK
model are found to agree with the new calculations.

\section*{Acknowledgements}

We would like to thank J. Bindslev Hansen, J. Kutchinsky and R. Taboryski
for useful discussions.

\widetext

\end{document}